**Thermally-driven atmospheric escape: Transition from hydrodynamic to Jeans escape**

Alexey N. Volkov[1], Robert E. Johnson[1,2], Orenthal J. Tucker[1], Justin T. Erwin[1]

[1]Materials Science and Engineering, University of Virginia, Charlottesville, VA 22904-4745

[2]Physics Department, NYU, NY, NY 10003-6621

**Abstract**

Thermally-driven atmospheric escape evolves from an organized outflow (hydrodynamic escape) to escape on a molecule by molecules basis (Jeans escape) with increasing Jeans parameter, the ratio of the gravitational to thermal energy of molecules in a planet's atmosphere. This transition is described here using the direct simulation Monte Carlo method for a single component spherically symmetric atmosphere. When the heating is predominantly below the lower boundary of the simulation region, $R_0$, and well below the exobase, this transition is shown to occur over a surprisingly narrow range of Jeans parameters evaluated at $R_0$: $\lambda_0 \sim$ 2-3. The Jeans parameter $\lambda_0 \sim 2.1$ roughly corresponds to the upper limit for isentropic, supersonic outflow and for $\lambda_0 > 3$ escape occurs on a molecule by molecule basis. For $\lambda_0 > \sim 6$, it is shown that the escape rate does not deviate significantly from the familiar Jeans rate evaluated at the nominal exobase, contrary to what has been suggested. Scaling by the Jeans parameter and the Knudsen number, escape calculations for Pluto and an early Earth's atmosphere are evaluated, and the results presented here can be applied to thermally-induced escape from a number of solar and extrasolar planetary bodies.

**Introduction**

Our understanding of atmospheric evolution is being enormously enhanced by extensive spacecraft and telescopic data on the outer solar system bodies and on exoplanets. The large amount of data on Titan's atmosphere from the Cassini spacecraft led to the use of models for atmospheric escape that predicted rates that differed enormously (Johnson 2009; Johnson et al. 2009). This disagreement was due to a lack of a kinetic model for how atmospheric escape changes in character from evaporation on a molecule by molecule basis to an organized flow, referred to as hydrodynamic escape, a process of considerable interest early in the early stages of the evolution of a planet's atmosphere (e.g., Watson et al. 1981; Hunten 1982; Tian et al. 2008) and to the evolution of exoplanet atmospheres (e.g., Murray-Clay et al. 2009).

The transition from hydrodynamic to Jeans escape is described in terms of the local Jeans parameter, $\lambda = U(r)/kT(r)$, where $U(r)$ is a molecule's gravitational energy at distance $r$ from the planet's center, $T$ is the local temperature, and $k$ is the Boltzmann constant (e.g., Chamberlain and Hunten 1987; Johnson et al. 2008). In Hunten (1982) it was suggested that if $\lambda$ decreases to ~2 above the exobase, then the escape rate is not too different from the Jeans rate, but if it becomes ~2 near or below the exobase, hydrodynamic escape can occur. Since it was assumed that the transition region from Jeans to hydrodynamic escape occurred over a broad range of $\lambda$, an intermediate model, referred to as the slow hydrodynamic escape (SHE) model, was developed to describe outflow from atmospheres such as Pluto's with exobase values $\lambda$~10 (Hunten and Watson 1982; McNutt 1989; Krasnopolsky 1999; Tian and Toon 2005; Strobel 2008a). In this model, based on that of Parker (1964) for the solar wind, the continuum fluid equations accounting for heat conduction are solved to an altitude where the flow velocity is still smaller than the speed of sound, and then asymptotic conditions on the temperature and density are applied. Unfortunately, this altitude is often above the nominal exobase. This model, subsequently applied to even Titan ($\lambda \sim 40$), could overestimate the mass loss rate (Tucker and Johnson 2009; Johnson 2010). Other recent models simply couple the Jeans rate to continuum models at the exobase, even when the escape rates are quite large (Chassefiere1996; Tian et al. 2009), or they couple to a modified Jeans rate (Yelle 2004; Tian et al. 2008). Here the transition from Jeans to hydrodynamic escape from a planetary body is described via a kinetic model.

Escape from an atmosphere can be calculated from the Boltzmann equation, which can describe both continuum flow and rarefied flow (e.g., Chapman and Cowling 1970) at large distances from a planet. Here we use the Direct Simulation Monte Carlo (DSMC) method (e.g. Bird 1994) to simulate thermally-driven flow in a one-dimensional radial atmosphere. For escape driven by heat deposited *below* the lower boundary of the simulation region, $R_0$, we show that the transition occurs over a surprisingly narrow range of Jeans parameters evaluated at $R_0$ ($\lambda_0 \sim 2 - 3$). That is, hypersonic outflow drives escape for $\lambda_0 \leq \sim 2$, but above $\lambda_0 \sim 3$ hypersonic flow *does not* occur, even at large distances from the exobase, and for $\lambda_0 > \sim 6$ the escape rate is close to the Jeans escape rate. Following the description of the DSMC model, results for a range of $\lambda_0$ are presented with emphasis on the transition region.

2. **DSMC simulations**

A kinetic model for calculations of the structure of the upper atmosphere and the escape rate is based on the Boltzmann kinetic equation. A DSMC method (Bird 1994), which is a stochastic method for numerical solution of problems based on the Boltzmann equation, is used here to simulate a spherically-symmetric, single component atmosphere supplied by out gassing from a surface at radius $R_0$. This can be the actual surface with a vapor pressure determined by the solar insolation or a virtual surface in the atmosphere, above which little additional heat is deposited and at which the density and temperature are known. In such a model, escape is driven by thermal conduction and heat flow from below $R_0$. In DSMC simulations, the real gas is simulated by means of large number of representative molecules of mass m. Trajectories of these molecules are calculated in a gravity field and subject to mutual collisions. It is readily shown that for a velocity independent cross section, the Boltzmann equations, and, hence, the results presented here, can be scaled by two parameters: the *source* values of the Jeans parameter, $\lambda_0$, and a Knudsen number, $Kn_0 = l_0/R_0$ where $l_0 = (2^{1/2} n_0 \sigma)^{-1}$ is the mean free path of molecules at the lower boundary with $\sigma$ the collision cross section and $n_0$ the number density on lower boundary. The Knudsen number often discussed for a planet's atmosphere is $Kn(r) = l/H$, where $l$ and $H$ are the local mean free path of molecules and the atmospheric scale height at radial distance $r$. Since $Kn(R_0) = \lambda_0 Kn_0$, it could be used, instead of $Kn_0$, as one of the two scaling parameters.

Simulations were conducted on a non-homogeneous mesh and the number of simulated molecules in a cell was varied over a wide range (e.g., Volkov et al. 2010). Results are presented for the hard sphere collisions, but comparisons made using variable hard spheres and forward directed collision models (e.g.,Tucker and Johnson 2009) gave similar results. In the kinetic model, the flow at each $r$ is described in terms of a velocity distribution function, $f(r,v_\parallel,v_\perp,t)$, where $v_\parallel$ and $v_\perp$ are velocity components parallel and perpendicular to radial direction and $t$ is time. Macroscopic parameters are calculated for molecular quantities $\Psi(r,v_\parallel,v_\perp)$ using the integral operator $\langle \Psi \rangle = 2\pi \int_{-\infty}^{+\infty}\int_0^\infty \Psi(r,v_\parallel,v_\perp) f(r,v_\parallel,v_\perp,t) v_\perp dv_\perp dv_\parallel$, e.g., number density, $n$ ($\Psi = 1$), radial flow velocity, $u$ ($\Psi = v_\parallel/n$), parallel, $T_\parallel$ ($\Psi = m(v_\parallel - u)^2/(nk)$), and perpendicular, $T_\perp$ ($\Psi = mv_\perp^2/(2nk)$), temperatures with $T = (T_\parallel + 2T_\perp)/3$. The molecules have gravitational energy $U(r) = -GMm/r$, where $M$ is the planet's mass and $G$ is the gravitational constant. The mass of the gas above $R_0$ is assumed to be much smaller than $M$ so that self-gravity is neglected. The velocity distribution in the boundary cell at $r=R_0$ is maintained to be Maxwellian at a fixed $n_0$ and $T_0$, and zero gas velocity. The exit boundary at $r = R_1$ is placed far enough from $R_0$ so the gas flow above this boundary can be approximately treated as collisionless. A molecule crossing $R_1$ with velocities $v_\parallel$ and $v_\perp$ will escape if $v > (-2U(R_1)/m)^{1/2}$, where $v = \sqrt{v_\parallel^2 + v_\perp^2}$, while a molecule with a smaller $v$ will return to $R_1$ with $-v_\parallel$ and $v_\perp$. Since collisions can modify the flow even at relatively large $r$, the effect of the position of $R_1$ was studied (Volkov et al. 2010) and $R_1$ is chosen to be sufficiently large in order to eliminate the effect of upper boundary in the flow region described. The simulations were carried out using the following fixed parameters, $m$, $\sigma$, $T_0$ and $R_0$ with $\lambda_0$ and $Kn_0$ varied by changing $M$ and $n_0$. However, the results scale with $\lambda_0$ and $Kn_0$ in a sense that any two flows with different $m$, $\sigma$, $T_0$, $R_0$, $M$, and $n_0$ represent the same flow in a dimensionless form, if the $\lambda_0$ and $Kn_0$ are the same.

Simulations are initiated by 'evaporation' from the cell at $R_0$ until steady flow is reached at large $t$. Steady-state flow is assumed to have occurred when the number flux $4\pi r^2 n(r)u(r) = \Phi$ varies by less that ~1% across the domain, where $\Phi$ is the escape rate. The Jeans escape rate, $\Phi_{Jeans}$, is defined by the upward flux of molecules with speeds $v \geq (-2U(r)/m)^{1/2}$ at the nominal exobase $r_{exo}$, typically determined by the scale height [$l(r_{exo}) = H(r_{exo})$] at large $\lambda_0$ or by the curvature [$l(r_{exo}) = r_{exo}$] at small $\lambda_0$. Due to the limited number of simulation particles, accurately representing the tail of escaping molecules in $f(r, v_\parallel, v_\perp, t)$ is problematic at large $\lambda_0$. For objects in the solar system, $\lambda_0$ varies from ~0.01 for comets to ~ 10 for escape from Pluto, and ~ 40 for escape from Titan, with much larger values for giant planet atmospheres. Since the total number of simulated molecules leaving $R_0$ in our simulations is typically ~$5 \times 10^8$, the simulations give accurate escape rates for $\lambda_0 < 15$ (Volkov et al. 2010). As will be seen, this is more than adequate for exploring the transition region. For $\lambda_0 \leq 10$ all simulations described below were performed with $R_1/R_0 = 40$, while simulations at $\lambda_0 = 15$ were performed with $R_1/R_0 = 6$, which is enough for accurate calculation of the escape rates. Distributions of gas parameters shown in Fig. 1 are found in simulations with $R_1/R_0 = 100$.

## 3. Results for thermal escape

A systematic study of thermal escape was performed for values of $\lambda_0$ from 0 to 15 and a large range of $Kn_0$ (Volkov et al. 2010), but here we describe results for small $Kn_0$, so that the lower boundary is assumed to be well into the collision dominated region. The atmospheric properties vs. $r$ for a number of $\lambda_0$ with $Kn_0 = 0.001$ are shown in Fig. 1. It is seen that the dependence on $r$ of the atmosphere density, the Mach number, and temperature are similar for the small $\lambda_0$ (0, 1, 2) cases in Fig. 1, but the nature of the radial dependence changes dramatically as $\lambda_0$ is increased (3, 10) with a distinct transition in the narrow range between 2 and 3 as indicated by the $\lambda_0 = 2.5$ case. In particular, at a given $r/R_0$, $n/n_0$, and $T_\perp/T_\parallel$ tend to increase with increasing $\lambda_0$ for $\lambda_0 \leq 2$ and tend to decrease with increasing $\lambda_0$ for $\lambda_0 \geq 3$. It is interesting that $T_\perp/T_\parallel$, which is considered a measure of translational non-equilibrium in expansion from a spherical source (e.g., Cattolica et al. 1974), remains close to unity in the transition region throughout most of the simulation domain as seen for $\lambda_0 = 2.5$. The degree of translational non-equilibrium, however, increases rapidly as $\lambda_0$ exceeds 3. Although the perpendicular component of the temperature is found to decay to zero, as expected, at large $r$, it is seen in Fig. 1c that $T_\parallel$ does *not* go to zero for the larger $\lambda_0$. More important to the escape problem of interest to us, the flow in Fig. 1b is seen to be *hypersonic* for the small values of $\lambda_0$, but *never becomes hypersonic* for the larger $\lambda_0$ shown, even at very large $r$ well above the nominal exobase. These aspects are contrary to the assumptions that are often made in the use continuum models for escape, such as the SHE model.

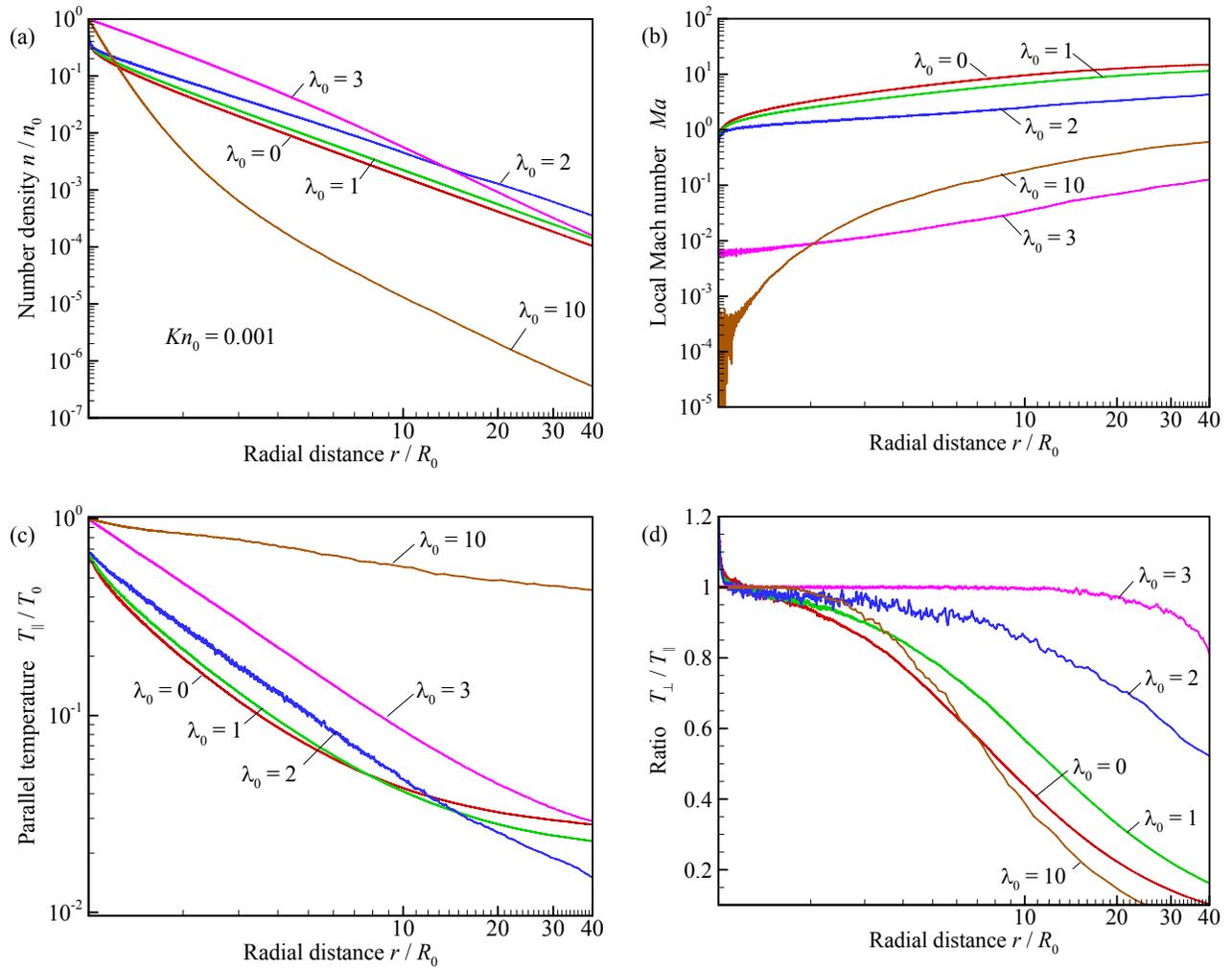

Fig. 1. Dimensionless number density $n/n_0$ (a), local Mach number $Ma = u/\sqrt{(5/3)kT/m}$ (b), parallel temperature $T_\parallel/T_0$ (c), and ratio of perpendicular to parallel temperature, $T_\perp/T_\parallel$, (d) vs. dimensionless

radial distance $r/R_0$ at $Kn_0$=0.001: $\lambda_0 = 0$ (red curve), 1 (green curve), 2 (blue curve), 3 (magenta curve), and 10 (brown curve). Curves for $\lambda_0 \leq 2$ are close to that for an isentropic expansion. Simulations are fully scaled by $\lambda_0$ and $Kn_0$: Actual simulations were performed for $N_2$ with fixed $m$=4.65x10$^{-26}$ kg, $\sigma$=7.1x10$^{-15}$ cm$^2$ (collision cross-section at 90 K (Bird 1994)), $T_0$ = 100 K, and $R_0$ = 1000 km, while $M$ and $n_0$ were varied.

The atmospheric properties at very small $Kn_0$ for $\lambda_0 \leq 2$ are found to roughly agree with those for an isentropic expansion of a monatomic gas above its sonic point $r_*$:

$$r^2 nu = r_*^2 n_* u_*; \quad \frac{T}{n^{2/3}} = \frac{T_*}{n_*^{2/3}}; \quad \frac{5}{2}kT + \frac{mu^2}{2} - \frac{GMm}{r} = \frac{5}{2}kT_* + \frac{mu_*^2}{2} - \frac{GMm}{r_*} \quad (1)$$

In Eq. 1, $n_*$, $u_*$, and $T_*$ are evaluated at $r = r_*$. For these small $\lambda_0$, $r_*$ occurs close to $R_0$ and hydrodynamic outflow occurs at small $Kn_0$. As $\lambda_0$ increases from zero to ~2, the flow gradually decelerates (Fig. 1b). In this range of $\lambda_0$, the thickness of the Knudsen layer decreases with decrease in $Kn_0$, $r_* \rightarrow R_0$ (Fig. 1c), and $T_* \rightarrow$ ~0.62$T_0$ (Fig 1d), the latter is consistent with kinetic simulations of spherical expansion at zero gravity based on the Bhatnagar-Gross-Krook equation (e.g., Sone and Sugimoto 1993). Therefore, it is seen that $[(5.2)kT_* + mu_*^2/2 - GMm/r_*]/kT_0 \rightarrow 0$ as $\lambda_0 \rightarrow$ ~2.1. This leads to rapid deceleration of the flow as $\lambda_0$ increases above 2.1. As a result, $n/n_0$ and $T_\parallel/T_0$ rise up to 1 on the source surface while the local Mach number $Ma$ rapidly drops below 0.1. The rapid transition from hydrodynamic outflow to a nearly isothermal atmosphere below the exobase is such that above $\lambda_0$~2.1 the sonic point is *not reached* at small $Kn_0$ up to very large distances from the source and the atmosphere is gravitationally dominated. Therefore, the transition region is *not* broad and $\lambda_0$~10 is well beyond the region where escape by outflow can occur, unlike what has been assumed in some models for Pluto's atmosphere (e.g., Hunten &Watson 1982; McNutt 1989; Krasnopolsky 1999; Strobel 2008a) and even Titan (Strobel 2008b; 2009).

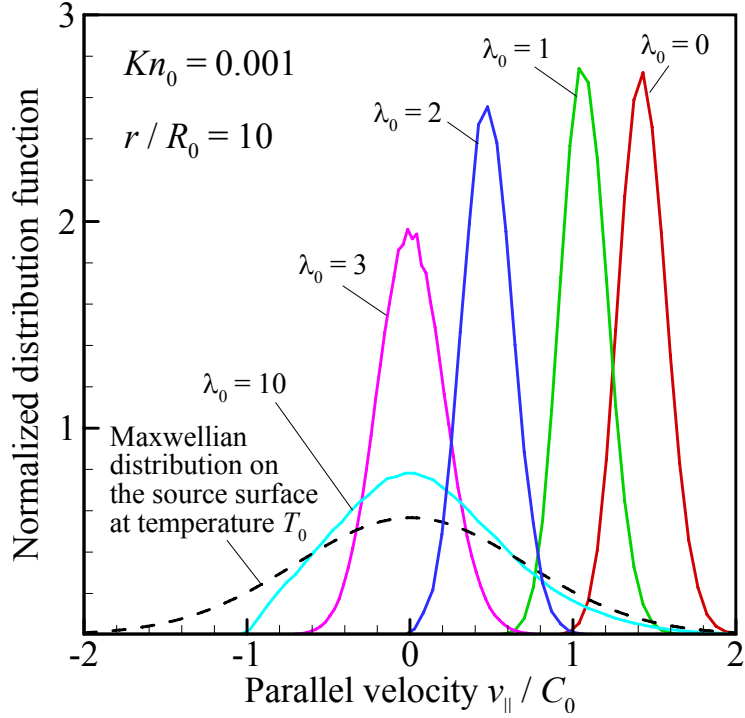

Fig. 2. Normalized distribution functions of parallel velocity $v_\parallel / C_0$ of gas molecules for $\lambda_0 = 0$ (red curve), 1 (green curve), 2 (blue curve), 3 (magenta curve), and 10 (cyan curve) at $Kn_0 = 0.001$ and $r/R_0 = 10$. Black dashed curves shows the Maxwellian distribution on the lower boundary at temperature $T_0$. $C_0 = \sqrt{2kT_0/m}$.

In the transition region the flow velocity, $u$, falls close to zero at large $r/R_0$. Above the transition region, the flow again accelerates at large $r/R_0$ as only fast molecules escape. This analysis is confirmed by looking at the distributions of the parallel component of molecular velocity, $v_\parallel$, in Fig. 2. For $\lambda_0 \leq 2$ the distribution is shifted at $r/R_0 = 10$ towards large $v_\parallel$ and the gas velocity, $u$, is close to the most probable molecular velocity, while at $\lambda_0 \geq 3$ the most probable $v_\parallel$ is close to zero and escape is provided by the depletion of the upward moving, high-speed molecules.

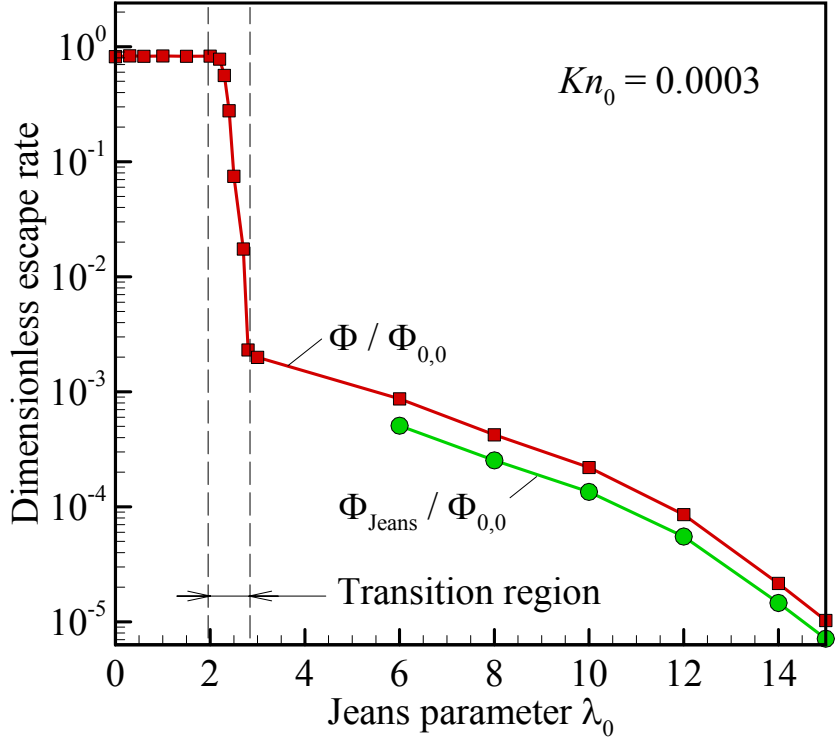

FIG. 3. Escape rate $\Phi/\Phi_{0,0}$ (red curve with square symbols) and the Jeans escape rate $\Phi_{Jeans}/\Phi_{0,0}$ evaluated at the nominal exobase (green curve with circle symbols) given as dimensionless ratio vs. Jeans parameter on the source surface $\lambda_0$ at $Kn_0 = 0.0003$. $\Phi_{0,0} = 4\pi R_0^2 n_0 \sqrt{kT_0/(2\pi m)}$ is the evaporation rate on the lower boundary. Vertical lines at $\lambda_0 = 2$ and $\lambda_0 = 2.8$ indicate the transition region.

These results have implications for the variation of the molecular escape rate, $\Phi$, with $\lambda_0$. In Fig. 3 we present a non-dimensional escape rate: the ratio of $\Phi$ to the escape rate for free molecular flow from the source surface, $\Phi_{0,0} = 4\pi R_0^2 n_0 \sqrt{kT_0/(2\pi m)}$. Remarkably, up to $\lambda_0 = 2$, this ratio is very close to ~0.82, the rate found in the absence of gravity as in a comet-like expansion (e.g., Cong and Bird 1978; Crifo et al. 2002; Tennishev et al. 2008). The ~0.82 is due to collisions in the Knudsen layer causing the return flow. As the atmospheric outflow is choked off by the increased gravitational binding, $\Phi/\Phi_{0,0}$ rapidly drops a couple of orders of magnitude between $\lambda_0 = 2$ to 3. In this transition region, escape is a product of a rapidly decreasing fraction of the velocity distribution function with $v > v_{esc}$ (e.g, Fig.2), assisted by a rapidly decreasing flow speed (e.g. Fig.1b). Therefore, for $\lambda_0 \geq 6$, it is seen that the escape rate is close to the Jeans rate. From Fig. 3 the ratio of the escape rate to the Jeans estimate varies from 1.7 to 1.4 as $\lambda_0$ goes from 6 to 15, consistent with earlier results (Tucker and Johnson 2009). A modified Jeans escape rate, accounting for the non-zero outflow gas velocity, $u$, can provide a good approximation for the actual escape rate at $\lambda_0 \leq 6$ as it is described in Volkov et al. (2010).

In scaling the escape rates using $\lambda_0$ and $Kn_0$, care must be taken. Unlike for the case $\lambda_0 \leq 2$, where $\Phi/\Phi_{0,0} \to \sim 0.82$ at $Kn_0 \to 0$, it is seen in Fig. 4 that escape rates at $\lambda_0 > 3$ for $Kn_0 = 0.0003$ cannot necessarily be consider a good approximation to the escape rate in the limit $Kn_0 \to 0$.

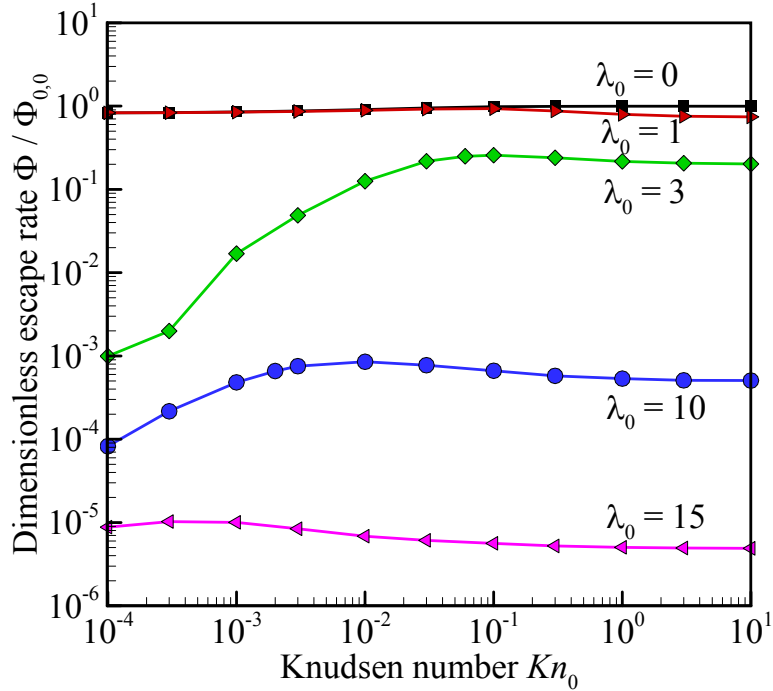

Fig. 4. Dimensionless escape rate $\Phi/\Phi_{0,0}$ v.s. Knudsen number $Kn_0$ at $\lambda_0 = 0$ (black curve), 1 (red curve), 3 (green curve), 10 (blue curve), and 15 (magenta curve). $\Phi_{0,0} = 4\pi R_0^2 n_0 \sqrt{kT_0/(2\pi m)}$ is the evaporation rate on the lower boundary.

In Fig. 5 DSMC results are also given for the thermal flux through the atmosphere for $\lambda_0$ = 10, a value relevant to Pluto's present atmosphere and considered an intermediate case between a comet outflow and a terrestrial-like atmosphere. It has been argued that, at such values of $\lambda_0$, escape driven by thermal conduction can be continued into the exobase region and above until the flow speed is some fraction of the speed of sound (e.g. Strobel 2008b). This picture has been shown to be incorrect (Johnson 2010). Here we note also that the flow speed does not reach the speed of sound at small $Kn_0$ (Fig. 1c), and from Fig. 5 it is seen that even a couple of scale heights below the exobase the heat flux is not well described by the Fourier law, $-\kappa(T)dT/dr$, where the thermal conductivity $\kappa(T)$ is defined by the first approximation of the Chapman-Enskog method (Chapman & Cowling 1970) and $T$ is the actual temperature found in the DSMC simulations. This law drastically overestimates the energy flux above an exobase defined by neutral-neutral collisions, consistent with $\lambda_0$ = 10 being well above the critical Jeans parameter.

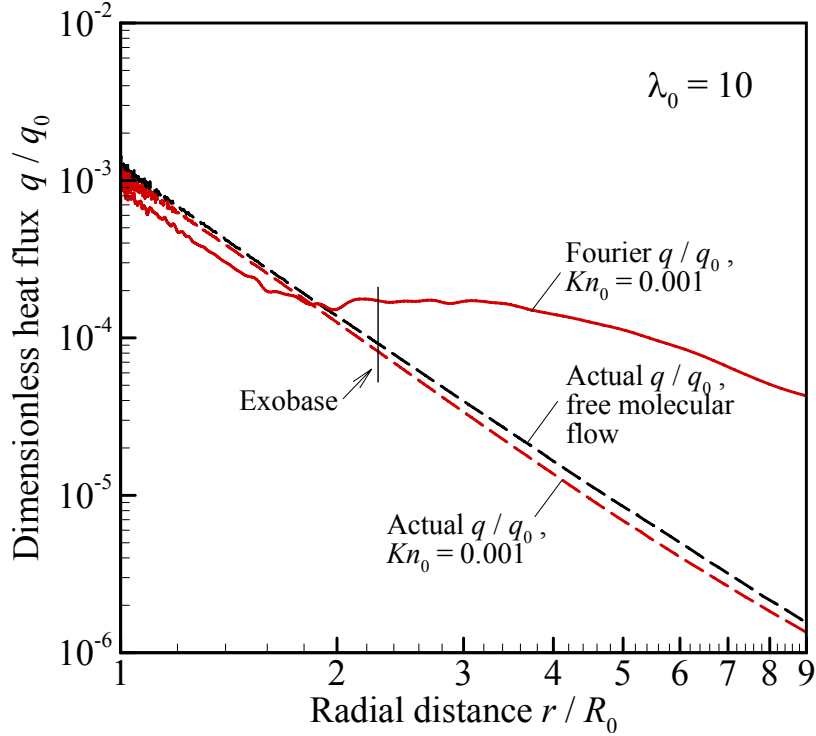

Fig. 5. Dimensionless heat flux $q/q_0$ vs. radial distance $r/R_0$ calculated for $\lambda_0 = 10$ at $Kn_0 \to \infty$ (free molecular flow: black dotted curve) and $Kn_0 = 0.001$ (red dashed curve). Solid curve represents dimensionless Fourier heat flux, where $q = -\kappa(T)dT/dr$ is calculated based on temperature $T(r)$ found in DSMC simulations at $Kn_0 = 0.001$: Thermal conductivity $\kappa(T)$ is given by the first approximation of the Chapman-Enskog method for a hard sphere gas (Chapman & Cowling 1970) and $q_0 = n_0 kT_0 \sqrt{2kT_0/m}$.

The results in Fig. 3 and 4 can now be used to evaluate previous calculations of atmospheric mass loss rates. For Pluto's predominantly nitrogen atmosphere, when all of the heating is assumed to be below $R_0 = 1450$ km, the Q=0 case in Strobel (2008a), then $\lambda_0 \sim 23$. From Fig. 3 the escape rate is seen to be very close to the Jeans rate and much smaller than $10^{-5}$ times the surface flux, $\Phi_{0,0}$. Therefore, the actual the mass loss rate is orders of magnitude below the rate estimated in that paper (~5x10$^{28}$ amu/s). Of more interest, is the mass loss rate from Pluto at solar medium heating: in Strobel (2008a) $Kn_0 = 0.01$ occurs at an $R_0$~3600 km, which is well above the solar heating maximum and corresponds to $n_0$ ~3.8x10$^7$ cm$^{-3}$ and $T_0 \sim$ 83K resulting in $\lambda_0 \sim 10$. Based on the results in Fig.4 the mass loss rate would be $\sim 2\times10^{27}$ amu/s. This is also more than an order of magnitude below that predicted: $\sim 9\times10^{28}$ amu/s. Therefore, it is clear that modeling in support of the New Horizon mission to Pluto will require iteratively coupling a kinetic model of escape to a fluid description of the lower atmosphere (Tucker et al 2010). As a second example, in a study of the EUV heating of the Earth's atmosphere, Tian et al. (2008) found the onset of hydrodynamic escape of oxygen, and the resulting adiabatic cooling of the thermosphere, occurs at an exobase having a value of $\lambda \sim 5.3$. Since this altitude corresponds to a $Kn_0 \sim 0.2$, and is well above the heating peak, the inferred onset is in disagreement with the results presented here. Since, they used a modified Jeans rate, their overestimate of the escape rate is only about a factor of 4-5. What may be more important is the effect on their description of the atmospheric structure near the exobase. Therefore, it is clear that fluid calculations of atmospheric escape should be tested against the results presents here, which by scaling can be used as upper boundary conditions.

**Summary**

A kinetic model for planetary atmospheres was used here to study the transition from hydrodynamic escape to escape on a molecule by molecule basis. When heat is deposited primarily below the lower boundary of the simulation region, $R_0$, and the collisions can be described by a hard sphere model, then the results presented can be *fully scaled* by two parameters evaluated at $R_0$: the Jeans parameter, $\lambda_0$, and the Knudsen number, $Kn_0$ (or $Kn(R_0)$). In the collision dominated regime at small $Kn_0$, this transition is found to occur over a surprisingly narrow range of $\lambda_0$, unlike what has been assumed in some continuum models of thermal escape. That is, below a critical value of $\lambda_0$ (~2.1), hydrodynamic outflow occurs, and is roughly described by an isentropic expansion starting from the sonic surface. Above the transition regime, $\lambda_0 > \sim 3$, escape occurs on a molecule by molecule basis and for $\lambda_0 > \sim 6$ we show that the escape rates do not deviate enormously from the Jeans rate. Although the simulations described here are highly idealized (hard sphere cross section and single component, spherically symmetric atmosphere) the overall conclusions will apply to the simulation of more complex atmospheres which are in progress. Therefore, re-evaluation of a number of studies, such as those for the early terrestrial atmospheres and the atmosphere of Pluto, soon to be visited by the New Horizon spacecraft, must be carried out using kinetic models of the upper atmosphere.

**Acknowledgements**

This work was supported by grants from NASA's Planetary Atmospheres Program and the Cassini Data Analysis Program.